# Indications of the topological transport by the universal conductance fluctuations in the Bi$_2$Te$_2$Se microflakes

Zhaoguo Li,[1] Yuze Meng,[1] Jian Pan,[2] Taishi Chen,[1] Xiaochen Hong,[2] Shiyan Li,[2] Xuefeng Wang,[1] Fengqi Song,[1,*] Baigeng Wang[1,*]

[1] *National Laboratory of Solid State Microstructures, Nanjing University, Nanjing 210093, P. R. China*

[2] *Department of Physics, Fudan University, Shanghai 200433, P. R. China*

E-mail: songfengqi@nju.edu.cn, bgwang@nju.edu.cn

**Abstract:**

Universal conductance fluctuations (UCFs) are extracted in the magnetoresistance responses in the bulk-insulating Bi$_2$Te$_2$Se microflakes. Its two-dimensional character is demonstrated by the field-tilting magnetoresistance measurements. Its origin from the surface electrons is determined by the fact that the UCF amplitudes keep unchanged while applying an in-plane field to suppress the coherence of bulk electrons. After considering the ensemble average in a batch of micrometer-sized samples, the intrinsic UCF magnitudes of over 0.37 $e^2/h$ is obtained. This agrees with the theoretical prediction on topological surface states. All the evidence point to the successful observation of the UCF of topological surface states.

**Text:**

The quantum interference transport of topological insulators (TIs) has been arousing much interest[1-3] due to the free-of-scattering and spin-blocking properties of the surface carriers protected by the time reversal invariance, where, however, the pin-down of the transport of the topological surface state (TSS) is still questionable.[4-11] Universal conductance fluctuation[12-14] (UCF), as an important manifestation of mesoscopic electronic interference, have been noticed in TIs recently.[10, 15-29] Giant conductance fluctuation (CF) amplitudes of 200-500 times over the expected have been observed in the mm-sized crystals.[20] The UCFs are further identified in some microflakes and demonstrated to be from two-dimensional (2D) interference by the field-tilting magnetoconductance (MC) measurements[10] and confirmed soon.[25, 26, 28] However, the question still exists whether such 2D UCFs are originated from a TSS since a few critical issues have to be taken care. In the previous work on the 2D UCF,[10] the bulk electrons fall into a crossover region between 2D and three-dimensional (3D) interference. This leads to the question that the 2D transport may arise from the bulk electrons. Such suspect is further strengthened by the possible bulk-surface coupling, which merges all the electronic states to a single 2D states.[30, 31] Another critical concern is the UCF contribution from the topologically-trivial 2D electron gas (2DEG) due to the surface band-bending, which has been shown by both spectroscopic[32] and calculation approaches[33] in practical TI samples. Here, we tackle the questions by studying the UCF effect in many samples of the bulk insulating $Bi_2Te_2Se$ (BTS) microflakes. The sample's thickness are chosen to eliminate the bulk-surface coupling and the 3D effects. A novel in-plane field measurement is proposed to exclude the bulk 2D interference. The intrinsic UCF is successfully extracted, supporting the TSS origin of the observed UCFs.

The BTS single crystals are grown by a high-temperature sintering method.[10] Then, all the microflakes are exfoliated from the same mother crystal and deposited on the $SiO_2$/Si substrates. The Au electrodes are applied onto the microflakes by a standard lift-off technique. The thickness (*H*) of all samples are measured by an atomic force microscopy (AFM). The typical samples can be seen in the insets of **Figure 1(a)** and **Figure 2(a)**. The resistance of all samples have been measured in four-probe configurations as shown in the left inset of Fig. 1(a). All the magnetotransport measurements are carried out in the Quantum Design Physical Property Measurement Systems. The in-plane field tuning is performed in an Oxford vector rotate magnet system.

The UCF can be extracted from the magnetotransport data. Fig.1 shows the transport data of a typical microflake with the thickness (*H*) of 60 nm (sample S4). The temperature-dependent resistance (*R*-*T*) reveals the bulk insulating of the microflake[25, 34, 35] [Fig. 1(a)]. Fig. 1(b) shows the MC as a function of the magnetic field. We can see a MC peak around the zero field, which is from the weak antilocalization[9] (WAL). In the high field range, there are some CF patterns. After subtracting a polynomial background curve [the red curve in Fig. 1(b)], we can clearly observe the aperiodic CF patterns [*δG-B*, the blue curve in Fig. 1(b)]. Such CF patterns can be observed repeatedly at different temperatures as shown in Fig. 1(c).

Please see the bottom curves in Fig. 1(c) for the two CF patterns, one of which is measured during the up-sweeping and the other is measured during the down-sweeping of the field $B$ [the arrows in Fig. 1(c) indicates the field-sweeping directions]. Despite the fact that the time interval between the two measurements is longer than 20 hours, the two CF curves still coincide with each other. In addition, the nearly same fluctuation features in $\delta G$-$B$ curves measured at different temperatures confirm the retraceability of the CFs. Such irregular but repeatable CFs are attributed to the UCFs of mesoscopic transport.[36, 37] Fig. 1(d) shows the measured $\delta G$-$B$ curves at various $\theta$. We can find the CF peaks in the $\delta G$-$B$ curves shift towards the high-$B$ direction and their widths are monotonically broadened with the increasing $\theta$, as guided by the circle-marked lines. The circle-markers represent the expected maxima for a 2D interference system, given by $B_\perp = B\cos\theta$. This indicates the UCF is from a 2D electronic interference.

An important reservation appears that the bulk interference is able to provide a 2D interference (UCF) in case that the thickness is close or less than the dephasing length of the bulk electrons. Actually, we have observed a MC curve at $\theta = 90°$ where the bulk electrons are dominant in the MC response. According to the traditional WAL theory,[38] we can obtain the dephasing length of bulk of around 60 nm. It implies that the bulk state falls into a crossover regime between the 2D and 3D interference. Hence, more evidence is required to distinguish the origin of the 2D UCF. The in-plane field ($B_\parallel$) tuning is an effective tool to exclude such bulk quasi-2D interference. As shown in the inset of Fig. 2(d), the UCF signal is adopted from $\delta G$-$B$ $\perp$ curves, but $B_\parallel$ only suppresses the coherence of the bulk states and doesn't disturb the coherence of surface states (SSs) because the closed diffusive paths of the SS carriers do not contain the magnetic flux of $B_\parallel$. As we know, the amplitude of UCF $\delta G_{rms} \propto L_\phi^{(4-d)/2}$ when $L_\phi \ll L$, where $d$ is the dimensions of system. While applying a $B_\parallel$, the dephasing length of bulk $L_{\phi,B}$ will be suppressed following $1/L_{\phi,B}^2 = 1/L_{\phi 0,B}^2 + 2eB_\parallel/\hbar$, where $L_{\phi 0,B}$ is the bulk's dephasing length at $B_\parallel = 0$ T. This will leads to the decreasing $\delta G_{rms}$ if the 2D UCF is partially contributed by the quasi-2D interfering bulk electrons.

**Figure 2** shows the magnetotransport tuned by $B_\parallel$ in a 47 nm-thick sample (sample S10). The insulating bulk is evident by inspecting its $R$-$T$ curve [Fig. 2(a)]. Fig. 2(b) shows the $\delta G$-$B_\perp$ curves at different $B_\parallel$. In order to quantitatively analyze the UCF, we calculate the CF autocorrelation function, which is defined by[12, 14]

$$F(\Delta B) = \langle [\delta G(B) - \langle \delta G(B) \rangle][\delta G(B + \Delta B) - \langle \delta G(B + \Delta B) \rangle] \rangle \quad (1)$$

Then, we can obtain the root mean square of CFs by using $\delta G_{rms} = \sqrt{F(0)}$. The dephasing length $L_\phi$ can be extracted by using the relation $(L_\phi)^2 B_c \sim \frac{h}{e}$, where $B_c$ is the half width at half maximum of the CFs autocorrelation function. Fig. 2(c) shows the $\delta G_{rms}$ as a function of $B_\parallel$, where we can see $\delta G_{rms}$ is independent of $B_\parallel$ in our samples. $L_{\phi,B}$ is ~ 50 nm by analyzing the $R$-$B_\parallel$ data.[39, 40] It will be reduced to ~ 17 nm while $B_\parallel = 1$ T, corresponding to a strong suppression of $\delta G_{rms}$ if the 2D UCF contributed by the bulk electrons. This is contradictory to our results in Fig. 2(c). Therefore, we exclude the bulk origin of the 2D UCF, and reasonably attribute the origin of UCF to some SSs, possibly TSS or trivial 2DEG. In Fig. 2(d), $L_\phi$ is plotted against $B_∥$, where no significant dependence between $L_ϕ$ and $B_∥$ can be seen. This further confirms the 2D UCF is originated from SSs. The $L_ϕ$ extracted from the UCF therefore describes the coherence of SSs.

We have investigated the magnetotransport of 14 BTS microflakes in this work. The device parameters of all samples are listed in Tab. I. The bulk insulating behaviors are identified in all samples [Fig. 3(c) and Tab. I]. The $δG$-$B$ curves of several samples are shown in Fig. 3(a). Fig. 3(b) shows the low-field MC [$ΔG(B) = G(B) − B(0)$] curves, which are identified as the WAL response originated from the π Berry phase of TSS.[1, 2, 31, 41] The similar 2D UCF and 2D WAL are observed in all the samples. The magnitudes of the UCF features fluctuate in different samples with different dephasing lengths.

The topological nature of the UCF can be demonstrated here. It has been suggested that the topological origin of the SS can be studied by the amplitudes of the 2D UCF in TI samples.[12-15, 17, 19] When the sample size $L$ is less than the dephasing length $L_ϕ$, recent theory have obtained a UCF amplitude $δG_{rms} = (0.43 ∼ 0.54)\ e^2/h$ for Dirac fermions (i.e. TSS),[17, 19] while $δG_{rms} = 0.86\ e^2/h$ for a normal 2DEG.[12-14] This indicates that we can distinguish the TSS from the 2DEG by directly measuring $δG_{rms}$ of TIs. However, the condition $L < L_ϕ$ fails in the experiments so far.[10, 20-29] The sample dimensions are normally a few micrometers while the dephasing lengths are often an order smaller. One may see very small $δG_{rms}$ values of around 0.01 $e^2/h$[10, 23, 25, 29] in the experiments, which can't be directly compared to the theoretical predictions.

To obtain the intrinsic UCF amplitudes in our samples, we consider the classical self-averaging effect, which often occurs in some independent phase-coherence segments in the mesoscopic samples. Note that the energy averaging may also make influence on the experimental UCF amplitudes, however, the effect can be neglected because the thermal diffusion lengths are comparable to the dephasing lengths in our samples.[42] The classical self-averaging modifies the UCF amplitudes as[12-14] $\delta G_{rms} \simeq \beta \frac{\delta G_{rms}^{\square}}{\sqrt{N}} \cdot \frac{W}{L} \simeq \delta G_{rms}^{\square} \cdot \beta \frac{L_\phi W^{1/2}}{L^{3/2}}$, where β is a suppression factor which is related to the symmetry of system and $\beta = 1/2\sqrt{2}$ in this work,[13, 43] $N \simeq L \times W/L_\phi^2$ is the number of independent phase-coherence segments, $L$ and $W$ are the length and width of the microflake respectively. The sheet conductance of a microflake $G_\square = G \cdot \frac{L}{W}$ is also considered. $\delta G_{rms}$ and $L_\phi$ can be extracted from the measured $\delta G$-$B$ curves, $L$ and $W$ are identified by AFM. Then, applying this formula to our samples, we can obtain the intrinsic UCF amplitude $\delta G_{rms}^{\square}$ of a single phase-coherence segment. We prepare 14 samples which are exfoliated from the same BTS crystal. All the data have been measured using the similar configuration described above and processed after considering the ensemble average. The results are shown in **Fig. 3(d)**, where $\delta G_{rms}^{\square}$ is plotted against $W/L$. Theoretically, the UCF amplitude of TSSs with a 2D geometry can be written as[15]

$$\delta G_{rms} = \frac{e^2}{\pi^2 h} \left( 12 \sum_{n_x=1, n_y=-\infty}^{\infty} \left[ n_x^2 + 4\left(\frac{L}{W}\right)^2 n_y^2 \right]^{-2} \right)^{1/2} \quad (2)$$

Please see Fig. 3(d), the solid curve is the theoretical prediction according to Eq. (2). We can see the UCF amplitude is decreasing with the increasing $L/W$. It soon saturates to 0.37 $e^2/h$ while $L/W \gtrsim 1$. Since all our samples fall into the $L/W > 1$ regime, $\delta G_{rms}^{\square}$ seems independent of $L/W$. These experimental data seem a little spread, but it can be explained by considering the impurity concentrations that are different in these samples or some inuniformity of the electronic configurations. Moreover, the difference of the Fermi level also affects the $\delta G_{rms}^{\square}$.[43] Please note that the experimental data are evenly distributed on both sides of the theoretical curve of TSS, and all data points are far below the theoretical value of a trivial 2DEG (dashed line). This excludes the contribution of the topologically-trivial 2DEG and reveals that the experimental results agree with the theoretical tendency of TSS. This essentially suggests we have accessed the UCF of a real TSS. To our knowledge, it has not been previously reported to measure the intrinsic UCF amplitudes of TSS in the bulk-insulating TI samples.

In summary, the UCF and its physical origin have been investigated in the bulk-insulating BTS microflakes. The 2D UCF features are demonstrated by the field-tilting analysis. The in-plane field tuning further excludes the contribution of the bulk electrons. We also investigate the classical self-averaging of the BTS's UCF to obtain the intrinsic UCF amplitude of over 0.37 $e^2/h$. All the results suggest that the UCF is originated from the TSSs.

**Acknowledgments** We thank the National Key Projects for Basic Research of China (grant numbers: 2013CB922103, 2011CB922103, 2010CB923401), the National Natural Science Foundation of China (grant numbers: 11023002, 11134005,

60825402, 61176088, 11075076, 21173040), NSF of Jiangsu province (No. BK2011592, BK20130016, BK20130054), the PAPD project, the NCET project and the Fundamental Research Funds for the Central Universities for financially supporting the work. Helpful assistance from Yanfang Wei and Prof. Zhiqing Li in Tianjin University, Wei Ning, Zhe Qu, Li Pi and Yuheng Zhang in High Magnetic Field Laboratory of CAS are also acknowledged. Insightful discussions with Dr. Ion Garate at SherBrooke University and Chris van Haesendonck at K. U. Leuven are greatly acknowledged.

**Figure captions:**

**Figure 1.** The UCF of a BTS sample with $H = 60$ nm. (a) The temperature dependence of its resistance. The left inset shows the measurement configuration. The right inset shows its AFM image. (b) A typical MC curve at $T = 2$ K. The red curve is the polynomial fitting result. The blue curve is the CF curve after subtracting the polynomial background. (c) The $\delta G$-$B$ curves at various temperatures ($\theta = 0°$). The arrows indicate the field-sweeping direction in the MC curves at 2 K. (d) The $B$-tilting $\delta G$-$B$ data measured at 2 K. The black, red and blue circle-markers represent the expected shift of the maxima in magnetic field for a 2D system, given by $B_\perp = B\cos\theta$. For clarity, adjacent curves in (c) and (d) are displaced vertically.

**Figure 2.** Tuning the UCF by $B_\parallel$ in a sample with $H = 47$ nm. (a) The temperature dependence of its resistance. The inset is its optical micrograph. (b) The $\delta G$-$B_\perp$ curves at various $B_\parallel$ values. The adjacent curves are displaced vertically. (c) $B_\parallel$ dependence of $\delta G_{\mathrm{rms}}$. (d) $B_\parallel$ dependence of $L_\phi$. The inset shows the magnetic field configuration. The data in (b-d) are measured at $T = 1.5$ K.

**Figure 3.** The UCF (a) and WAL (b) features of several samples at $T = 2$ K, and their corresponding $R$-$T$ curves are shown in (c). The adjacent curves in (a) and (b) are vertically shifted for clarity. (d) Intrinsic UCF amplitudes as a function of $L/W$. The closed-circle markers present the experimental data and the solid curve shows the theoretically-expected values according to Eq. (2). The dashed line presents the expected values for a topologically-trivial 2DEG.

**Table I.** The basic parameters of devices. $L$ is the distance between the two

voltage probes in a four-probe configuration. $W$ and $H$ are the width and height (thickness) respectively. $\delta G_{\text{rms}}$ is the measured root mean square value of CFs at $T = 2$ K. The resistance ($R$) of samples at $T = 2$ K and 300 K are also shown.

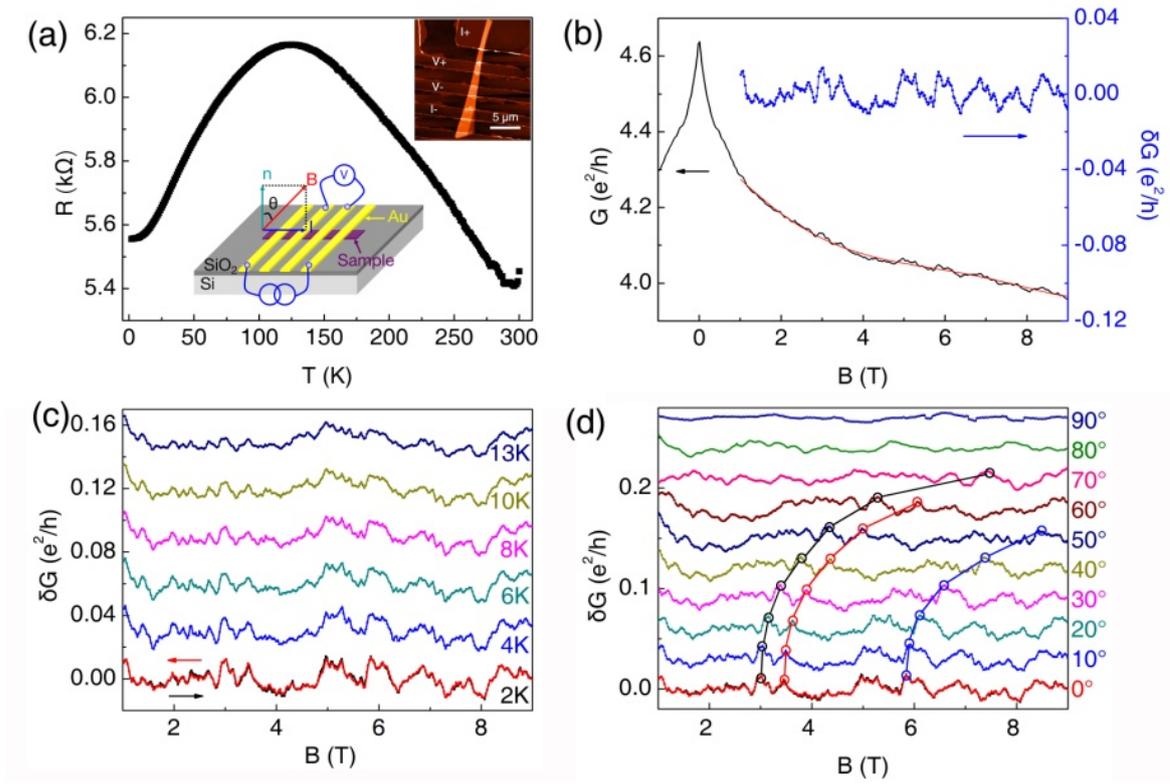

**Figure 1**

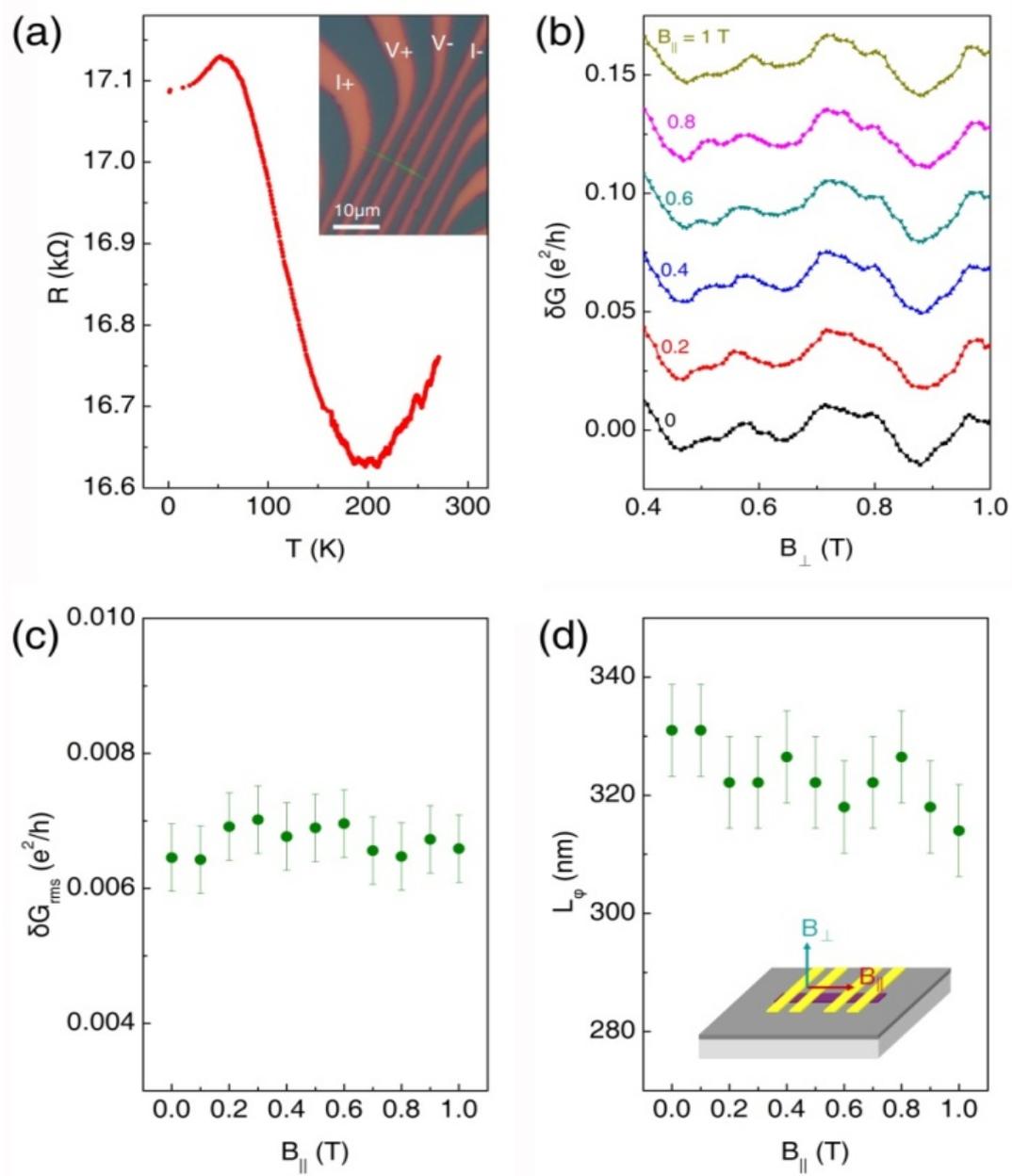

**Figure 2**

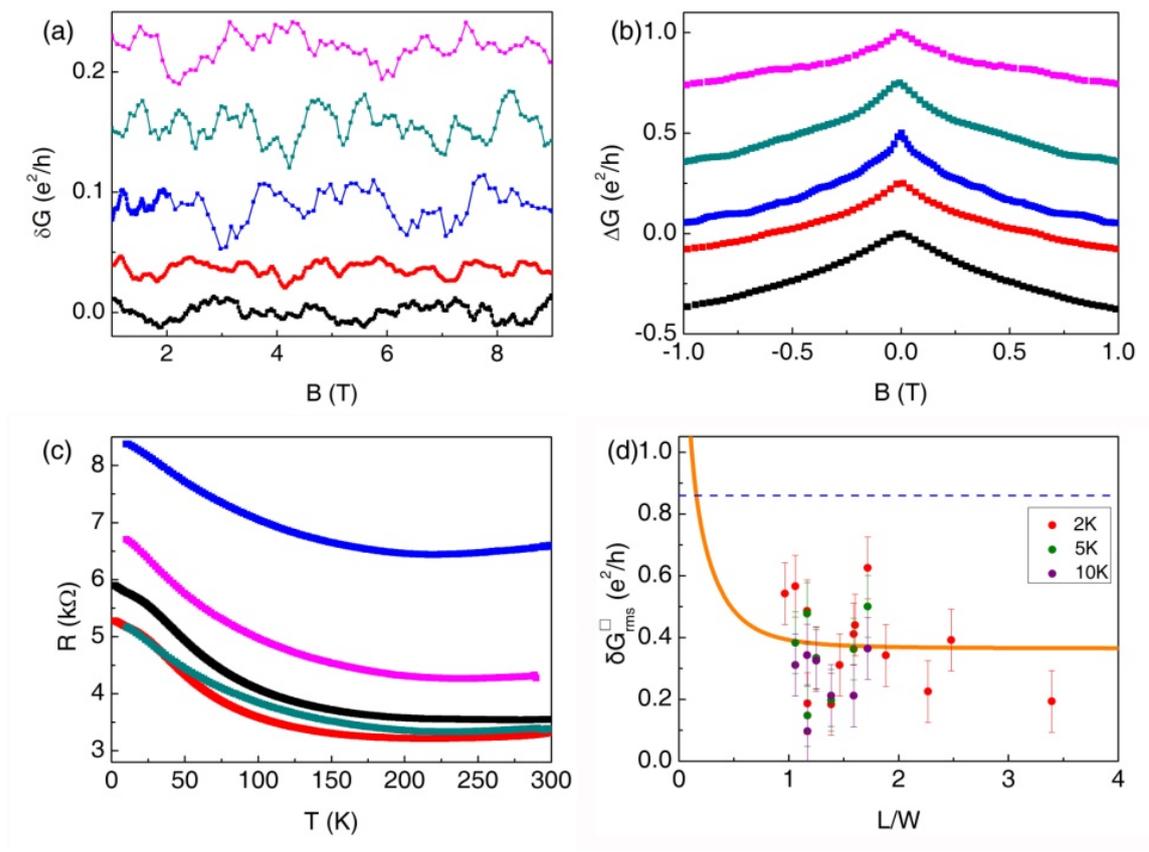

**Figure 3**

**Table I**

| Sample | L (μm) | W (μm) | H (nm) | R (2K) (kΩ) | R (300K) (kΩ) | $\delta G_{rms}$ ($e^2/h$) |
|---|---|---|---|---|---|---|
| S1 | 1.46 | 0.85 | 47 | 5.92 | 5.78 | 0.0149 |
| S2 | 1.40 | 1.20 | 50 | 4.07 | 3.66 | 0.0141 |
| S3 | 1.36 | 0.60 | 59 | 5.40 | 4.99 | 0.0074 |
| S4 | 1.66 | 1.20 | 60 | 5.55 | 5.45 | 0.0056 |
| S5 | 1.48 | 1.40 | 60 | 3.29 | 2.78 | 0.0201 |
| S6 | 1.80 | 1.54 | 61 | 5.90 | 3.55 | 0.0043 |
| S7 | 1.50 | 1.20 | 62 | 6.14 | 3.89 | 0.0082 |
| S8 | 1.40 | 0.88 | 98 | 7.86 | 6.67 | 0.0107 |
| S9 | 1.76 | 0.71 | 57 | 8.38 | 6.58 | 0.0070 |
| S10 | 1.90 | 0.56 | 47 | 17.09 | 16.76 | 0.0065 |
| S11 | 1.25 | 0.78 | 45 | 9.60 | 8.60 | 0.0148 |
| S12 | 1.53 | 1.59 | 70 | 7.62 | 6.41 | 0.0147 |
| S13 | 1.55 | 1.06 | 45 | 5.17 | 3.47 | 0.0117 |
| S14 | 1.79 | 0.95 | 58 | 6.70 | 4.28 | 0.0093 |